# Nanoscale Raman Characterization of a 2D Semiconductor Lateral Heterostructure Interface

*Sourav Garg[1,†], J. Pierce Fix[2,†], Andrey V. Krayev[3], Connor Flanery[2], Michael Colgrove[2],*

*Audrey R. Sulkanen[4], Minyuan Wang[4], Gang-Yu Liu[4], Nicholas J. Borys[2,‡], and Patrick Kung[1,§]*

[1]Department of Electrical and Computer Engineering, The University of Alabama,

Tuscaloosa, AL, 35487, USA

[2]Department of Physics, Montana State University, Bozeman, MT, 59717, USA

[3]HORIBA Scientific, Novato, CA, 94949, USA

[4]Department of Chemistry, University of California Davis, Davis, CA, 95616, USA

**Abstract:** The nature of the interface in lateral heterostructures of 2D monolayer semiconductors including its composition, size, and heterogeneity critically impacts the functionalities it engenders on the 2D system for next-generation optoelectronics. Here, we use tip-enhanced Raman scattering (TERS) to characterize the interface in a single-layer $MoS_2/WS_2$ lateral heterostructure with a spatial resolution of 50 nm. Resonant and non-resonant TERS spectroscopies reveal that the interface is alloyed with a size that varies over an order of magnitude—from 50-600 nm—within a single crystallite. Nanoscale imaging of the continuous interfacial evolution of the resonant and non-resonant Raman spectra enables the deconvolution of defect-activation, resonant enhancement, and material composition for several vibrational modes in single-layer $MoS_2$, $Mo_xW_{1-x}S_2$, and $WS_2$. The results demonstrate the capabilities of nanoscale TERS spectroscopy to elucidate macroscopic structure-property relationships in 2D materials and to characterize lateral interfaces of 2D systems on length scales that are imperative for devices.

† These authors contributed equally

‡ nicholas.borys@montana.edu

§ patkung@eng.ua.edu

Two-dimensional (2D) transition metal dichalcogenide (TMD) semiconductors provide a rich platform for the study of low-dimensional many-body phenomena[1] and for the development of next-generation optoelectronic and photonic technologies that exploit strong light-matter interactions in the atomically-thin limit.[2-4] Lateral 2D heterostructures[5-8] extend these capabilities with nanoscale interfaces between 2D materials with different stoichiometries, band structures, strain, and/or carrier densities. 2D heterostructures with *p-n* junction characteristics,[9] precisely engineered band alignment,[10] interfacial electroluminescence,[11] and efficient thermal transport[12] have been experimentally realized. The resulting interfaces provide an additional nanoscale knob for tailored exciton dissociation,[13] interfacial exciton formation,[14] carrier transport,[15, 16] photocurrent generation,[17, 18] polariton lenses,[19] and 1D charge-density waves.[19] With precision growth techniques, superstructures comprised of multiple transitions and interfaces can be embedded into a single 2D crystalline system, giving rise to even richer hierarchical heterostructures.[20, 21]

In order to fully harness the technological potential of 2D lateral heterostructures, it is crucial to understand the nanoscale structure and heterogeneity of the interface. High-resolution scanning transmission electron microscopy (STEM) measurements have revealed that the atomic structures of the interface range from atomically sharp[6] to more gradual alloyed[5] transitions. However, such atomically-resolved characterization has not elucidated how the interface changes on length scales larger than 10 nm and with different orientations, both of which are critical for understanding devices based on 2D lateral heterostructures. Scanning probe microscopy (SPM) techniques can bridge the atomically insightful length scales of STEM to the larger length scales that are needed for understanding device performance. Kelvin probe force microscopy (KPFM)[10, 22-24] has been used to map the built-in fields in *p-n* junctions and differences in work functions in lateral 2D heterostructures. Near-field scanning microwave microscopy (SMM)[25, 26] has been used to image local photoconductivity. While these techniques are powerful characterization approaches, they lack unambiguous stoichiometric sensitivity and probe a response that convolves the structure of the interface with non-local electrostatic properties. Nano-optical tip-enhanced photoluminescence (nano-PL) is sensitive to material composition and has been used to characterize nanoscale charge transfer processes[27] and the stoichiometry of larger junction regions.[28, 29] However, nano-PL techniques are limited to samples that exhibit strong PL emission, and the large linewidths of PL spectra limit their sensitivity to composition. Raman spectroscopy is capable of overcoming these two limitations, providing a characterization approach that is highly versatile and highly sensitive. Yet, to date, nanoscale Raman imaging and spectroscopy have not been used to probe the interfaces in 2D lateral heterostructures.

Here, we show that tip-enhanced Raman scattering (TERS) imaging and spectroscopy is a powerful and complementary characterization technique for heterostructure interfaces in 2D TMD semiconductors. Spatial resolutions of at least 50 nm are achieved, and we show that both resonant and non-resonant TERS measurements are sensitive to the local stoichiometry of the heterojunction. The TERS measurements are complemented with KPFM imaging of the heterostructure and are shown to work on a system where photoluminescence (PL) is strongly quenched by an underlying gold substrate. From the TERS characterization, we find that the transition region of the heterostructure drastically varies in spatial extent with sizes ranging from ~600 nm to less than 50 nm within the same crystallite. Using the nanoscale vibrational finger-printing, we assess the local alloying of the heterostructure and map the evolution of the Raman spectrum across the heterostructure interface. Insight on the origins and nature of multiple Raman modes is gained by mapping their nanoscale transformation across the interface. Overall, TERS is demonstrated as an effective way to study the nanoscale properties of heterostructure interfaces that complements other imaging techniques, laying a foundation for deeper multimodal studies that connect the interfacial structure and composition to macroscopic device performance.

**Results/Discussion**

**Characterization of the as-grown lateral heterostructure**

Single-layer (1L) lateral 2D heterostructures consisting of a 1L-$MoS_2$ core surrounded by a 1L-$WS_2$ shell were grown on polished sapphire wafers using low-pressure chemical vapor deposition (CVD; see Methods). As-grown crystals were first characterized using several SPM techniques that were cross-correlated with confocal PL imaging and spectroscopy. As discussed in detail in the Methods section, the TERS, SPM, and confocal characterization were performed with a commercial combined AFM and optical microscopy system (Xplora-Nano AFM-Raman, Horiba Scientific). SPM characterization of a representative lateral heterostructure crystal is summarized in Figure 1. Figure 1a shows a schematic of the lateral heterostructure which consists of a 1L-MoS$_2$ triangular core that is surrounded by a 1L-WS$_2$ shell. The optical image of the sample (Figure 1b) reveals that the typical shape of the 1L-MoS$_2$/1L-WS$_2$ lateral heterostructure crystallites is triangular with jagged edges, which are thought to arise due to growth conditions such as the chalcogen saturation and gas flow rate.[30] Additional optical imaging of the heterostructure crystallites are provided in the SI that demonstrate the jagged peripheral edges also form in the 1L-WS$_2$ shell (Figure S1 in the SI). The AFM topography of the lateral heterostructure (Figure 1c) shows that the transition between the 1L-MoS$_2$ and the 1L-WS$_2$ is flat, which confirms the predominantly lateral nature of the heterojunction. A small increase in the topography of the heterostructure of ~0.5 nm is observed around the periphery of the 1L-MoS$_2$ core. This topographic feature, which is smaller than the thickness of 1L-MoS$_2$ and 1L-WS$_2$, is attributed to a change in the interaction between the AFM probe and the 2D material as the composition changes (rather than as a sign of overlapping materials). As can be seen from both the topography and the contact potential difference (CPD; Figure 1d) maps, the interface between the 1L-MoS$_2$ core and the surrounding 1L-WS$_2$ is not a straight crystalline edge. Instead, the boundary of the 1L-MoS$_2$ core consists of small, predominantly, triangular protrusions. A capacitance map of the same area is shown in Figure 1e, which further confirms the contrast between the two materials of the single-layer heterostructure.

Figure 2 presents confocal μPL imaging and spectroscopy of the same 2D lateral heterostructure crystallite that is shown in Figure 1. Spatially-resolved PL emission was recorded over the wavelength range of 550 nm to 750 nm under CW laser excitation (532 nm, 300 μW, 25 ms/pixel integration). Figures 2a-2d map out the integrated intensity of the PL spectrum for the entire emission band (Fig. 2a), a high-energy band for the 1L-WS$_2$ (Fig. 2b), an intermediate band between those of the two pure materials (Fig. 2c), and a low-energy band for the 1L-MoS$_2$ (Figure 2d). Bright PL emission from all regions of the 2D lateral heterostructure confirms the monolayer nature of the 1L-MoS₂ core and surrounding 1L-WS₂ as shown in the full spectra in Figure 2e. In the 1L-MoS₂ core, the PL is centered at ~670 nm (1.85 eV), which is expected for the emission of 1L-MoS₂. In the surrounding 1L-WS₂, the PL is ~6× brighter than the 1L-MoS₂ core and is centered at ~628 nm (1.97 eV), as expected for 1L-WS₂. In the interfacial region between the 1L-MoS₂ and 1L-WS₂, the PL intensity increases compared to the 1L-MoS₂ core and its spectrum is positioned at an intermediate wavelength of 638 nm (1.94 eV). As a result, a peripheral ring of bright emission in the intermediate band is formed at the heterostructure interface (Fig. 2c).

The spectrum of the intermediate PL in the heterojunction region indicates that the interface is comprised of a ternary alloy of 1L-Mo$_x$W$_{1-x}$S₂.[31-33] The average relative Mo and W content in this alloy region can be estimated from the spectral position using Vegard's Law that describes the change in the PL energy with alloy stoichiometry:

$$E_{Mo_xW_{1-x}}(x) = (x)E_{MoS_2} + (1-x)E_{WS_2} - b(x-1)x.$$

Here, the $b$ is the bowing factor, which has been reported as 0.25 eV[31] for TMD alloys. Using this relationship and the emission energies here, the composition would be coarsely estimated to be 1L-Mo$_{0.09}$W$_{0.91}$S₂ in the interfacial region (*i.e.*, $x \approx 0.09$). However, the linewidth of the PL spectrum in the interfacial region (~130 meV, FWHM) is substantially larger than those of the 1L-MoS₂ (~50 meV, FWHM) and the 1L-WS₂ (~65 meV), suggesting the presence of disorder. As shown below, the interfacial disorder is dominated by a continuous range of alloy compositions and transition widths. The heterostructure is a more complex optoelectronic system than an atomically sharp interface or just a simple uniform alloy composition, necessitating characterization with higher spatial resolutions.

Disentangling nanoscale phenomena at the heterostructure interface necessitates a different approach than conventional SPM and confocal optical techniques. The diffraction-limited resolution of the μPL (and μRaman) characterization limits the ability to directly probe structure-property relationships of the 2D lateral heterostructure interface at length scales smaller than a few hundred nanometers. For SPM techniques that have a higher spatial resolution, unambiguous signatures of material alloying and interfacial disorder are not available. Bridging this divide, nano-optical characterization such as nano-PL,[28, 29, 34-39] nano-absorption,[40] and TERS (*i.e.*, nano-Raman)[36, 41, 42] enables materials characterization with nanoscale spatial resolutions, especially for 2D material systems. Nano-optical techniques utilize a nanoscale plasmonic antenna to nano-focus optical fields to sub-wavelength volumes which can then be used to study highly localized light-matter interactions.[43] In particular, TERS enables vibrational fingerprinting of materials at the nanoscale. Sub-nanometer spatial resolutions have been achieved in ultrahigh vacuum (UHV) with STM modalities.[44-47] Under ambient conditions, AFM-based TERS can routinely achieve resolutions of 10-20 nm,[36, 48] offering a 50× improvement over μRaman. Moreover, 2D materials are particularly amenable to gap-mode TERS techniques (Figure 3a) that enhance the signal-to-noise ratio by positioning the nano-optical antenna in close proximity to a smooth metal film (<5 nm) to create a nanoscale plasmonic cavity that enhances the local electric field even further. With the ability to provide high signal-to-noise nanoscale vibrational fingerprinting, gap-mode TERS is a compelling technique to probe the nanoscale properties of interfaces in lateral 2D heterostructures in a manner that synergizes the strengths of inelastic optical spectroscopy (*e.g.*, Figure 2) with the spatial resolution of SPM techniques (*e.g.*, Figure 1).

**Semi-resonant TERS imaging and spectroscopy**

Gap-mode TERS characterization requires the 2D system to be supported by a smooth metallic substrate. To implement the gap-mode TERS characterization here, the 2D crystals were transferred to a gold film using a previously established procedure[49] (see Methods). A thin gold film (~70 nm) was deposited onto the surface of the as-grown 2D crystallites on the sapphire growth substrate under high vacuum. Then, a Si wafer was attached to the outer gold surface using an epoxy resin. Once the epoxy cured, the gold-sapphire interface was separated by peeling. The 2D lateral heterostructures are more strongly bound to the Au film and are thus separated from the sapphire substrate. The process results in the transfer of the 2D lateral heterostructures from the sapphire surface to being inlaid in the gold film,[49, 50] exposing the pristine surfaces of the 2D lateral heterostructures that were previously in contact with the sapphire substrate.

An optical image of the 2D lateral heterostructure that was characterized in Figures 1 and 2 after the gold-assisted transfer process is shown in Figure 3b. Using optical contrast microscopy, the same single-layer crystallites were easily identified before and after the gold-assisted transfer process (see Figure 2 in the SI). As can be seen by the AFM characterization in Figures 3c and 3d, the transfer process preserves the smooth topography of the 2D lateral heterostructure. The variation in the topography in Figure 3c is less than 0.5 nm indicating that tearing or fracturing did not occur during the transfer process. Further KPFM characterization of the 2D lateral heterostructure partially embedded in the gold shows clear contrast in the CPD between the 1L-$MoS_2$ core and the surrounding 1L-$WS_2$ shell (Figures 3e and 3f), which reflects the differences in work functions between the two materials. The difference in the CPD between the 1L-$MoS_2$ and 1L-$WS_2$ is ~150 mV, consistent with earlier reports.[23, 24] The CPD imaging on the Au surface (Fig. 3e) more clearly reveals that the apparent boundary between the 1L-$WS_2$ and 1L-$MoS_2$ is intricately profiled, exhibiting sharp saw-tooth-like protrusions of 1L-$MoS_2$ into the surrounding 1L-WS$_2$. The extent of the transition between the 1L-MoS$_2$ and 1L-WS$_2$ in terms of the CPD is 500-1000 nm, which, as shown below, is significantly larger than what is measured using TERS. The likely origin of this difference is the sensitivity of CPD measurements to the work function of the material, which depends on both the stoichiometry and the depletion region of the heterostructure. As a result, CPD measurements will report a transition width that is larger than the size of the material interface as determined by just stoichiometry (see Table S1 for a summary of prior CPD measurements of 2D heterostructure which all report similar size scales)[51-53].

Gap-mode TERS characterization of the 2D lateral heterostructure provides deeper insight into the nature of the interface. The TERS imaging and spectral analysis using laser excitation at 638 nm are presented in Figure 4. To facilitate direct comparison, Figure 4a replicates the region of the CPD imaging in Figure 3e for the region that was characterized with gap-mode TERS. We note that this region does not have any characteristics in the SPM imaging that distinguish it from the other transition regions in the crystallite. The laser excitation at 638 nm is weakly resonant with both the long-wavelength tail of absorption for the 1S exciton state in 1L-WS$_2$ and the short-wavelength tail of the absorption of the 1S exciton state of the 1L-MoS$_2$. Under these semi-resonant conditions, the TERS spectra (cf. Figure 4b) from the 1L-MoS$_2$ core and the 1L-WS$_2$ shell (*i.e.*, away from the heterojunction interface) exhibit many features that are similar to resonant μRaman spectroscopy[54, 55] but have notable differences that highlight the potential for gap-mode TERS to augment insight gained from far-field μRaman measurements. For the 1L-WS$_2$ shell, the prominent mode at 417 cm$^{-1}$ corresponds to the $A'(\Gamma)$ mode, whereas the cluster of modes from 296 cm$^{-1}$ to 355 cm$^{-1}$ reflects the resonant nature of the laser excitation.[55] Complementing these prominent modes, four low-energy vibrational modes are identified at 148, 176, 200, and 215 cm$^{-1}$. Previous studies have reported three of these four modes (146.5, 176, and 214 cm$^{-1}$) and have assigned them to the ZA($M$), E$'(M)^{\text{TO2}} - $LA($M$) or LA($M$), and E$''(M)^{\text{TO1}} - $TA($M$) modes, respectively[56-58]. However, the relative intensities of these peaks relative to, for instance, the $A'(\Gamma)$ mode are larger in gap-mode TERS (compared to those observed in µRaman on a dielectric interface).[55-58] In addition, the gap-mode TERS spectrum of the 1L-WS$_2$ also exhibits a mode at 433 cm$^{-1}$, which to our knowledge has not been observed in far-field µRaman measurements. Prior combined scanning tunneling microscopy (STM) and TERS measurements of 1L-WS$_2$ have correlated this mode to S vacancies,[59] which as discussed below, agrees with our observations here.

The gap-mode TERS spectrum of the 1L-MoS$_2$ (Figure 4b, blue) also reflects the semi-resonant nature of the laser excitation at 638 nm. The most prominent feature is the vibrational mode observed at 455 cm$^{-1}$. While this mode is known to emerge under resonant conditions, its assignment has not been unambiguously established. It could arise from the combination of 2LA($M$) and the normally IR active $A_2''/A_{2u}(\Gamma)$ modes or from the higher energy component E$''(M)^{\text{TO1}} + $ZA($M$) mode.[54, 57] As will be discussed below, the TERS imaging here also directly links this mode to the 433 cm$^{-1}$ mode in the 1L-WS$_2$, suggesting that its origin is also related to S vacancies. At the lower energies, multiple modes are observed between 320 cm$^{-1}$ to 440 cm$^{-1}$. For non-resonant Raman scattering, two modes are anticipated in this spectral region. The first mode corresponds to $E'(\Gamma)$ at 386 cm$^{-1}$ and $A'(\Gamma)$ at 405 cm$^{-1}$. Prior reports[54, 57, 60] of resonant Raman scattering spectra of 1L-MoS$_2$ on dielectric substrates do not observe modes between 386 cm$^{-1}$ and 405 cm$^{-1}$. On gold, it has been shown[61] that charge transfer splits the $A'(\Gamma)$ mode, producing a second mode at 400 cm$^{-1}$ which was also accompanied by a ~7 cm$^{-1}$ shift of the $E'(\Gamma)$ mode to lower wavenumbers. Here, we observe the intermediate mode at 400 cm$^{-1}$ but do not observe the shift of the $E'(\Gamma)$ mode. These differences that are observed in the resonant Raman spectra of 1L-WS₂ and 1L-MoS₂ between conventional μRaman and gap-mode TERS could be due to a number of potential factors such as less strain in our system, a stronger out-of-plane polarization in gap-mode TERS as well as subtle interactions with the tip (*i.e.*, induced strain[38] or charge transfer[62]). Identifying the underlying mechanisms is outside the scope of the current work and will be the subject of future investigations that will focus on a careful comparison of polarization-resolved μRaman to gap-mode TERS.

Figures 4c and 4d report the integrated peak intensity for the most prominent modes that are exclusive to 1L-MoS₂ and 1L-WS₂, respectively. To improve the contrast, these images report the peak intensity which is calculated by integrating the intensity of the spectrum in a specific band after subtraction of a linear background. The peak-filtering analysis provides an alternative for extracting peak intensities to multi-peak fitting routines, which necessitate relatively high signal to noise ratios, especially when dealing with overlapping peaks (see Section 4 in the SI for details). A transition region between the 1L-MoS₂ core and the 1L-WS₂ shell is identified and its width dramatically varies between a pixel-limited size to a few hundred nanometers at the tips of the triangular protrusions of 1L-MoS₂. The gap-mode TERS spectrum acquired from this transition region (Figure 4b, orange) is most similar to that of the 1L-WS₂ shell in terms of the spectral positions and relative intensities of the major peaks but has substantial distinguishing features. The first obvious difference is that a pronounced mode at 444 cm$^{-1}$ is observed and lies between the positions of the 455 cm$^{-1}$ mode of the 1L-MoS₂ and the 433 cm$^{-1}$ S-vacancy defect mode of the 1L-WS₂. Second, the intensities of the 200 cm$^{-1}$ and 215 cm$^{-1}$ modes are substantially higher for the transition region as compared to that of 1L-WS₂. The increase in the intensity of these modes suggests that they can be used to image the transition region directly with a large amount of contrast. Figure 4e shows the integrated peak intensity of these modes over the lateral 2D heterostructures and demonstrates that these modes are the strongest in the transition region where the prominent modes for 1L-MoS$_2$ and 1L-WS$_2$ are the weakest. It also confirms the size variations of the transition region separating the core and shell of the lateral 2D heterostructure. By using independent color channels to render the intensities for the 1L-MoS$_2$ (red), 1L-WS$_2$ (green), and transition region (blue), a full-color image can be generated that shows how these different regions align with one another, depicting the nanoscale structural configuration of the transition region (Fig. 4f).

Considering the μPL characterization (*i.e.*, Figure 2), the distinct Raman bands in the transition region are potential nanoscale reporters of alloy formation between the 1L-MoS$_2$ and 1L-WS$_2$ regions. To test this hypothesis, we conducted two control measurements on known alloyed samples (see Sections 5 and 6 in the SI for details). In the first control, we carried out TERS characterization of CVD-grown 1L-Mo$_x$W$_{1-x}$S$_2$ crystallites where the growth produces a spatial gradient of composition. In the second, we performed gap-mode TERS spectroscopy of a series of known 1L-Mo$_x$W$_{1-x}$S$_2$ alloys that were mechanically exfoliated from commercially available bulk crystals. These measurements confirmed that the appearance and intensification of the modes between 140 cm$^{-1}$ and 220 cm$^{-1}$ under resonant excitation at 632.8 nm are associated with the alloying in Mo$_x$W$_{1-x}$S$_2$ compounds. These modes are practically absent in pure 1L-MoS$_2$ and 1L-WS$_2$ and reach maximum intensity in the alloyed compounds tested when the tungsten to molybdenum ratio is 2.32. Because these modes are observed throughout the 1L-WS$_2$ region of the 2D lateral heterostructure, these controls indicate that there is a persistent alloying in the shell region away from the junction, which is further supported by the TERS imaging of the transition region reported below (see also Figure S7 in the SI). In terms of the mode in the transition region at 444 cm$^{-1}$, no such mode was observed in the control measurements on single layers exfoliated from bulk crystals. Additionally, in the pure 1L-WS$_2$ that was mechanically exfoliated, the 433 cm$^{-1}$ mode is not observed, which is consistent with the expectation that the bulk crystals have fewer chalcogen vacancies than the CVD-grown crystallites. Finally, the linewidths of the PL spectra of the control alloys are smaller than that of the transition region, again indicating the presence of interfacial disorder as previously discussed.

From the TERS imaging presented in Figure 4, the width of the alloyed transition between the 1L-WS$_2$ shell and 1L-MoS$_2$ core varies dramatically, ranging from the resolution limit of the measurement (*i.e.*, <50 nm) to ~600 nm. Figure 5 reports the progression of the TERS spectra across a representative transition region that is broad (~600 nm in length; Fig. 5a) and one that is sharp (<50 nm in length; Fig. 5b). Over the broad transition region, the TERS spectrum continuously transforms from that of 1L-MoS$_2$ to alloyed 1L-Mo$_x$W$_{1-x}$S$_2$ to 1L-WS$_2$. This evolution provides insight into the relationships of the modes across the different materials. The 455 cm$^{-1}$ mode of the 1L-MoS$_2$ continuously shifts to lower wavenumbers in the alloyed region and ultimately converges to the 433 cm$^{-1}$ mode in the 1L-WS$_2$, which directly connects these respective modes in 1L-MoS$_2$ and 1L-Mo$_x$W$_{1-x}$S$_2$ to the mode in 1L-WS$_2$ that is activated by S vacancies. Similarly, the cluster of modes between 400-420 cm$^{-1}$ in the 1L-MoS$_2$ evolves into the 417 cm$^{-1}$ mode in the 1L-WS$_2$. In contrast to a continuous shift, the 1L-MoS$_2$ modes in the range of 380-400 cm$^{-1}$ gradually weaken over the transition region, whereas the 1L-WS$_2$ mode at ~360 cm$^{-1}$ intensifies over the transition region nearer to the 1L-WS$_2$ shell as the composition becomes W-rich. The low-energy modes (140–220 cm$^{-1}$) become brightest halfway across the transition region. Because these modes are resonantly enhanced, this onset marks the alloy composition at which the A exciton state is maximally resonant with the excitation laser, providing direct evidence that the heterostructure interface is composed of the full range of alloy compositions.

The two representative transition regions shown in Figure 5 confirm that the size of the heterostructure interface can vary by at least an order of magnitude on the same 2D heterostructure crystal. Due to the resolution-limited size of the sharp transition region (Fig. 5b), the transition from the 1L-MoS$_2$ to the 1L-WS$_2$ is discrete and does not exhibit further changes beyond the interface of the two materials. Given prior high-resolution TEM characterization and the alloyed nature of the broad transition, the sharp transition is likely similarly alloyed at length scales below our current resolution. Strikingly, the heterogeneity in the size of these transition regions occurs on microscopic length scales within the same 2D crystallite. The distance between these two regions is ~400 nm, essentially eliminating the possibility that the different concentrations of reactants during growth as the origin of the heterogeneity. Rather, we hypothesize that the origin of the heterogeneity may lie in the different reactivities of edges with different orientations.

### Non-resonant TERS imaging and spectroscopy

To further probe the transition region and the nature of the Raman modes observed under semi-resonant excitation, we performed TERS characterization with non-resonant laser excitation at 785 nm (Figure 6), which is lower in energy than the optical bandgaps of both 1L-WS$_2$ and 1L-MoS$_2$. Under these conditions, the TERS spectra of the 1L-WS$_2$, 1L-MoS$_2$, and transition regions are substantially different than their counterparts under semi-resonant excitation (cf. Figures 6a, 6b, and 4b). The resonant and non-resonant TERS measurements were conducted on the same region of the sample to facilitate direct comparison. The topography of this region is reported in the SI (see Figure S6) and does not exhibit any topographic features that correspond to the transition. With the non-resonant excitation, the lower-energy Raman modes in the 140-220 cm$^{-1}$ range are no longer observed anywhere in the 2D heterostructure. Further, the intensity of the $E'(\Gamma)$ peak at ~353 cm$^{-1}$ is greatly reduced in the 1L-WS$_2$ shell and the transition region. Instead, the non-resonant Raman spectrum is dominated by the lower-energy compositional-dependent $A'(\Gamma)$ mode which evolves from 405 cm$^{-1}$ in the 1L-MoS$_2$ core to 420 cm$^{-1}$ in the 1L-WS$_2$ shell. Comparable behavior for this mode has been observed in μRaman spectroscopy of 1L-Mo$_x$W$_{1-x}$S$_2$ alloys, spanning a similar range of energies.[57, 63] At higher energies (450-460 cm$^{-1}$), the presumed defect mode remains active under non-resonant laser excitation and evolves in the opposite direction as the $A'(\Gamma)$ mode. It decreases in energy from 455 cm$^{-1}$ in the 1L-MoS$_2$ core to 435 cm$^{-1}$ in the 1L-WS$_2$ shell. In the transition region, this mode bridges the two energies and appears at 442 cm$^{-1}$, and as shown in the SI (see Figure S4) and below, depends on the particular alloy composition. By mapping the peak intensity of this mode in high-energy (436-456 cm$^{-1}$), intermediate-energy (430-450 cm$^{-1}$), and low-energy (420-440 cm$^{-1}$) bands around this mode, the 1L-WS$_2$ shell, transition, and 1L-MoS$_2$ core regions can be clearly discerned (Figures 6c-6f) and agree with the same type of imaging under resonant excitation (Figure 4).

The differences in the Raman spectra between resonant and non-resonant laser excitation provide insight into the nature of the observed modes. When the laser excitation energy corresponds to an electronic transition in the material, resonant enhancement of weak/forbidden Raman modes can occur.[64] Furthermore, in gap-mode TERS, the polarization of the electric field is predominantly normal to the surface (*i.e.*, out-of-plane). In such a configuration, it was recently shown on small molecular systems that non-resonant excitation preferentially enhances out-plane modes, whereas the resonant excitation can dramatically and preferentially enhance in-plane modes.[45] The suppressed intensity of the $E'(\Gamma)$ mode (at ~386 cm$^{-1}$ in the 1L-MoS$_2$) under non-resonant conditions suggests that similar selectivity in enhancement occurs in 2D TMDs as the in-plane mode is substantially brighter under resonant excitation. Likewise, the practical absence of the rarely reported peaks within the 140-220 cm$^{-1}$ range demonstrates that they are forbidden (*i.e.*, very weak) modes under non-resonant conditions. Further, the resonant enhancement of these modes explains their observed dependence on alloy concentration as discussed above. The resonance energy of the 1S transition of the A exciton evolves from 1.85 eV in pure 1L-MoS$_2$ to 2.0 eV in pure 1L-WS$_2$ following Vegard's law (cf. discussion above) and is maximally resonant with the 633 nm excitation at an alloy concentration of 1L-Mo$_{0.23}$W$_{0.77}$S$_2$. This optimal composition agrees well with the TERS spectra reported in the SI (See Figure S5) for different alloy compositions, where the maximum intensity of these modes is determined to occur at an alloy composition between 1L-Mo$_{0.15}$W$_{0.85}$S$_2$ and 1L-Mo$_{0.30}$W$_{0.70}$S$_2$.

Finally, in Figure 7, the evolution of the non-resonant TERS spectra from the 1L-MoS$_2$ core to the 1L-WS$_2$ is analyzed. As under the resonant excitation conditions, transition regions that are pixel-limited in size are identified alongside those that are over 500 nm in size, confirming the heterogeneity of the 2D lateral heterostructure interface. Figures 7a and 7b trace the evolution of the TERS spectra across representative broad (~500 nm) and sharp (~100 nm) transition regions, respectively. As under resonant excitation conditions, the modes under non-resonant excitation continuously evolve from the 1L-MoS$_2$ to the 1L-WS$_2$ across the broad region (a line in Fig. 7a inset). Both of the 405 cm$^{-1}$ and the 455 cm$^{-1}$ modes in the 1L-MoS$_2$ shift into the 420 cm$^{-1}$ and 435 cm$^{-1}$ modes in the 1L-WS$_2$, respectively. In the transition region, each of these modes appears at intermediate energies between the respective extremes. As the energies of these modes depend on the alloy composition, this observation confirms the alloyed nature of the transition region. Beyond the transition region in the MoS$_2$ core and the WS$_2$ shell, the energies of these modes do not evolve further (see Figure S7 in the SI). Future multimodal measurements that correlate these TERS signatures with, for example, nano-Auger imaging[34] could further confirm these observations and provide additional clarity into the origin of the many modes observed in the TERS spectrum as well as the heterogeneous nature of the transition region at length scales below 50 nm.

## Conclusions

In conclusion, TERS is a facile and informative characterization technique for heterostructure interfaces in lateral heterostructures between 2D materials. Here, the interface between a lateral heterostructure of 1L-MoS$_2$ and 1L-WS$_2$ has been characterized with a synergistic combination of resonant TERS, non-resonant TERS, confocal μPL, AFM, and KPFM imaging techniques. In particular, the TERS and other scanning probe techniques provide critical nanoscale information that bridges the gap between atomically-resolved STEM studies and diffraction-limited μPL/μRaman studies. On these intermediate length scales at which TERS is ideally suited to probe, we find that the transition region can be highly heterogeneous in size, ranging from widths of <50 nm to 600 nm in a single crystallite. We hypothesize that the different sizes of the transition region result from different reactivities of the edges that form during the growth process. Future studies that use scanning tunneling microscopy and/or nano-second harmonic generation imaging to probe the chemical nature of the edges could test this hypothesis by providing more insight into the structure and reactivity of the edges on the nanoscale. Further, for the TERS characterization, the contributions of resonant excitation and alloy composition have been deconvolved for low-energy vibrational modes in the range of 140-240 cm$^{-1}$. By mapping the continuous nanoscale evolution of the transition region from 1L-MoS$_2$, to 1L-Mo$_x$W$_{1-x}$S$_2$, to 1L-WS$_2$, we unambiguously link the 455 cm$^{-1}$ mode in 1L-MoS$_2$ to the 433 cm$^{-1}$ defect-activated mode in 1L-WS$_2$ and reveal how the mode changes with alloy composition. This mode serves as an excellent nanoscale reporter of the composition of the transition region and is used to characterize the heterogeneity in the size of the interface in the 2D heterostructure. These studies provide important insight into the nanoscale properties of the heterostructure interfaces in 2D materials and set the stage for combining TERS characterization with other microscopy modalities and prototype optoelectronic devices to quantitatively understand the role of the heterostructure interface in device performance.

### Methods/Experimental

**Heterostructure growth:** The single-layer MoS$_2$/WS$_2$ lateral heterostructures were grown by low-pressure chemical vapor deposition in a multizone quartz tube. High purity MoS$_2$ and WS$_2$ precursor powders (both 99.9%, Alfa Aesar) were finely mixed and placed in a quartz boat at the center of the tube while the epi-ready sapphire substrates were placed downstream at a lower temperature. No prior substrate surface treatment or preparation was used beyond solvent cleaning. The growth was performed at a pressure of 10 mbar under 20 sccm argon flow, with the temperature of the precursors ramped to 970 °C, for a duration of 20 min. The tube temperature for the substrate region was ramped from 790 to 810 °C during the growth. Subsequently, the temperature was allowed to cool down naturally.

**Optical, scanning probe, and TERS characterization:** The AFM and TERS characterization presented in the main manuscript were conducted on an XploRA-Nano AFM-Raman system (HORIBA Scientific) using Access-SNC-Au TERS probes (Applied Nanostructures). TERS maps were collected using patented SpecTop$^{TM}$ mode, where, in every pixel of the map, the tip is placed in direct contact with the sample and the Raman data are collected. Then, to move to the next pixel, the system is rapidly placed into semicontact (tapping) mode to minimize the ware of both the tip and the sample. Measurements of the TERS response of the gold-transferred exfoliated $Mo_xW_{(1-x)}S_2$ crystals presented in the SI were performed on a TRIOS AFM-Raman platform (HORIBA Scientific) coupled to an imaging spectrograph (Andor) and an electron multiplying charged coupled device (EM-CCD). There TERS measurements of the exfoliated samples were performed using 632.8 nm continuous wave excitation and incident laser intensity of 50 µW. For all TERS spectra, a constant background is subtracted to account for the dark counts of the detectors and any broad emission. The dark counts are estimated from a constant baseline intensity at low wavenumber spanning from 100-400 $cm^{-1}$. This background is relatively constant over the course of the TERS measurements and did not exhibit strong pixel-to-pixel or row-to-row variations.

**Transfer of 2D crystallites onto gold thin films:** To perform gap mode TERS measurements, the crystals needed to be on a gold substrate. A thin gold film (~70 nm) was deposited onto the sample under high vacuum ($1.0 \times 10^{-7}$ Torr) using a thermo evaporator (model DV502-A, Denton Vacuum Inc., Moorestown, NJ). Then, a Si wafer was attached to the outer gold surface using an epoxy resin. Once the epoxy cured, the gold film and sapphire substrate were separated by peeling, transferring the 2D lateral heterostructures to the gold surface. Figure S2 in the SI provides evidence of the transfer of the 2D crystallites from sapphire to the gold surface upon peeling.

**Supporting Information Available:** additional data and discussions that describe the results of the gold transfer process, the peak filtering analysis, the control spectroscopy measurements, a comparison of prior characterization results, and the experimental techniques are presented in the Supporting Information, which is available online.

**Acknowledgments**

N.J.B. acknowledges support from the Murdock Charitable Trust through award SR-201811596 and the NSF Q-AMASE-i program (Award# 1906383). This work was performed in part at the Montana Nanotechnology Facility, a member of the National Nanotechnology Coordinated Infrastructure (NNCI), which is supported by the National Science Foundation (Grant# ECCS-2025391). Support from the University of Alabama ORED SGP is also acknowledged.

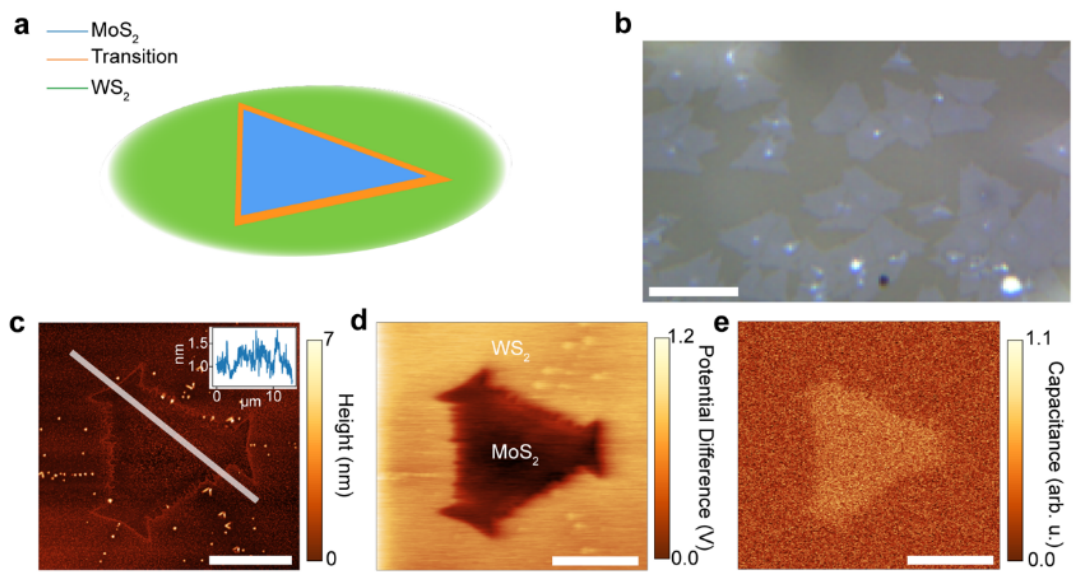

**Figure 1** – Scanning-probe characterization of the as-grown 2D lateral heterostructure of 1L-MoS$_2$ and 1L-WS$_2$. (a) Schematic of the 2D lateral heterostructure which is composed of a core of 1L-MoS$_2$ that is separated from a shell of 1L-WS$_2$ by a transition region composed of a 1L-Mo$_x$W$_{1-x}$S$_2$ alloy of varying width. (b) Optical image of the as-grown lateral heterostructures. Scale bar: 50 μm. (c) AFM topography, (d) CPD, and (e) capacitance images of the single lateral heterostructure crystallite. Inset in (c): topographic profile of the lateral heterostructure along the path denoted by the white line in (c). Scale bars in (c), (d), and (e): 5 μm.

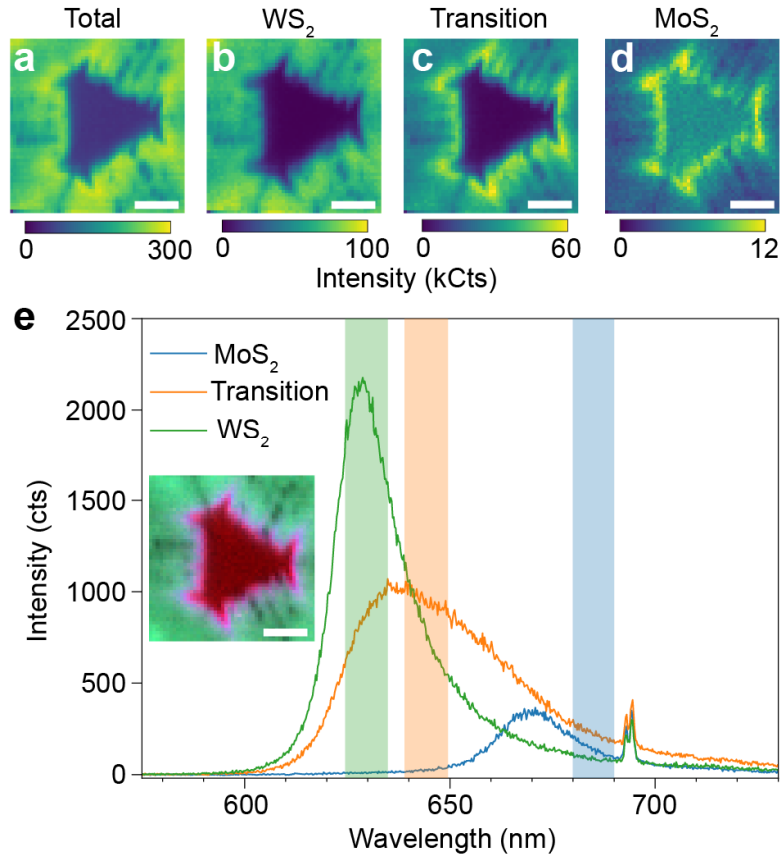

**Figure 2** – Confocal μPL imaging and spectroscopy of the as-grown 2D lateral heterostructure of 1L-MoS₂ and 1L-WS₂. Spatial maps of (a) the total emission intensity, (b) a 10 nm emission band for 1L-WS₂, (c) a 10 nm intermediate band for the transition region, and (d) a 10 nm emission band for 1L-MoS₂. (e) Example spectra for the 1L-MoS₂ core, 1L-WS₂ shell, and the transition region. The shaded regions indicate the bands used to generate the images in (b)-(d). The narrow emission lines between 690-770 nm are ruby emission from the sapphire substrate. Inset: combined images of the 1L-MoS₂, 1L-WS₂, and the transition regions rendered in the red, green, and blue channels respectively. All scale bars: 4 μm.

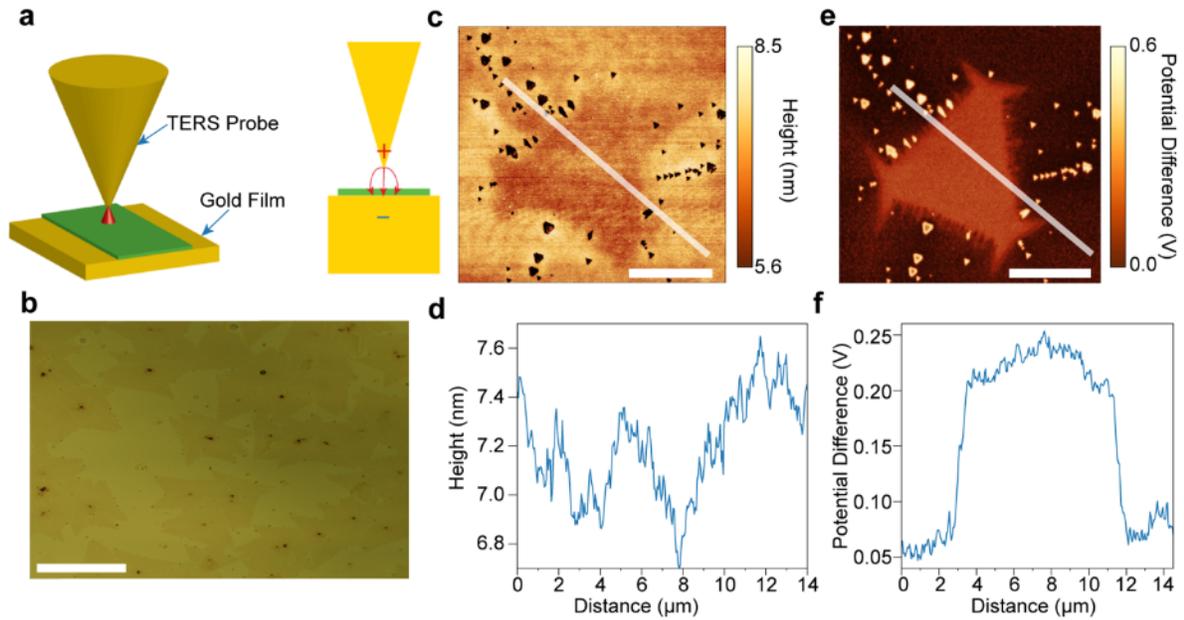

**Figure 3** – Hybrid Au-2D lateral heterostructure system for gap-mode TERS characterization. (a) Schematic of gap-mode TERS where the 2D material is stripped from its growth substrate using an Au-assisted stripping technique. As a result of the template stripping process, the lateral heterostructure is embedded in the Au. (b) Optical image of a stripped 2D 1L-MoS$_2$/1L-WS$_2$ lateral heterostructure. Scale bar: 50 µm. (c) AFM image of the lateral heterostructure embedded in the Au. Scale bar: 5 µm. (d) Topographic profile of the 2D heterostructure along the white line in (c). (e) CPD image of the lateral heterostructure embedded in Au. Scale bar: 5 µm. (f) CPD profile along the white line in (e), which corresponds to the topographic profile in (c) and (d).

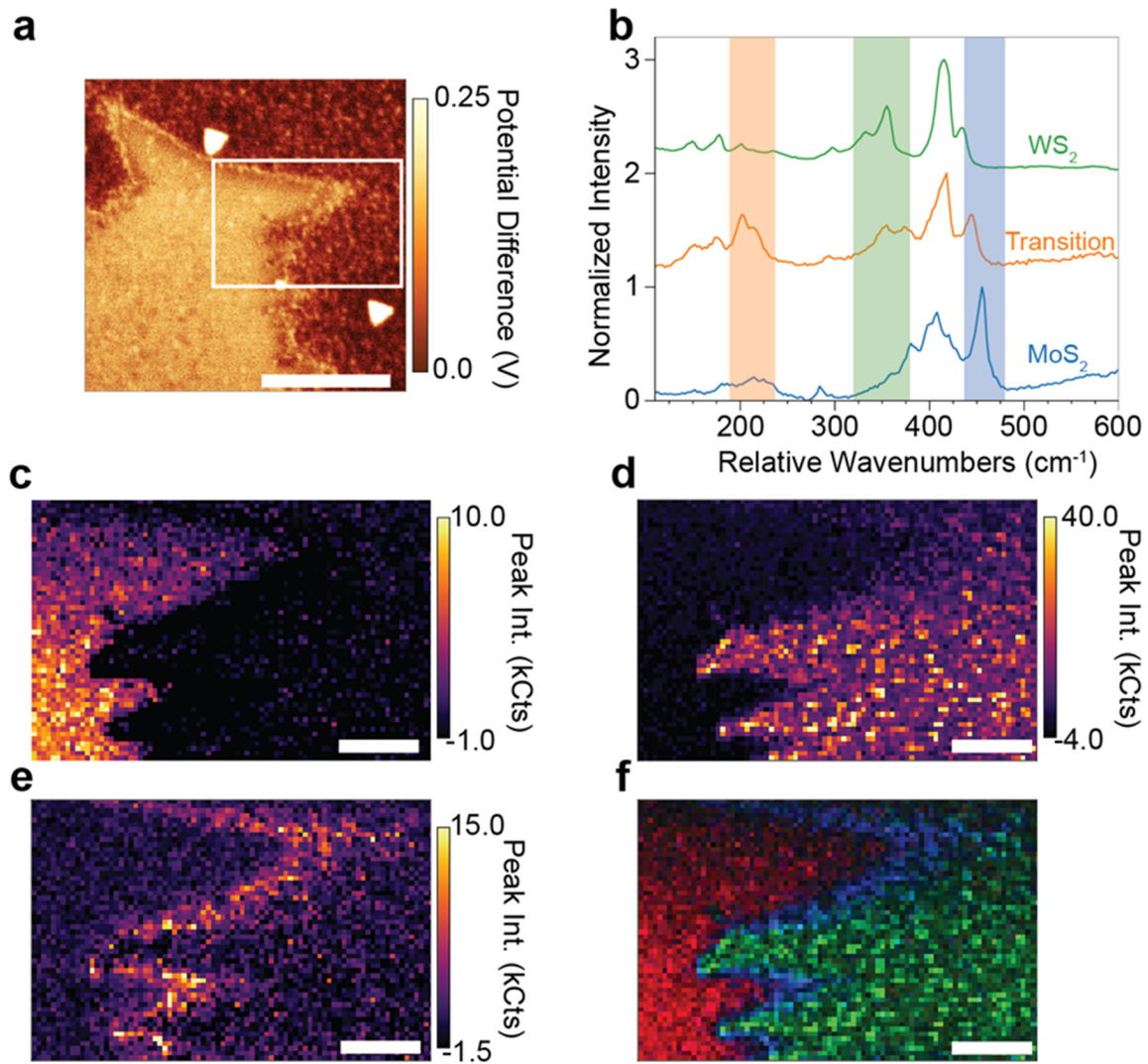

**Figure 4** - semi-resonant TERS imaging and spectroscopy of the heterostructure interface. (a) CPD image of the region characterized with TERS. Scale bar: 2 μm. (b) Representative semi-resonant TERS spectra of the 1L-MoS₂ core, 1L-WS₂ shell, and transition regions of the lateral heterostructure. The spectra are averages over the following areas: 350 pixels from a 370×840 nm² area for MoS$_2$, 42 pixels from a 222×168 nm² for transition, and 400 pixels from a 740×270 nm² for WS$_2$. A constant background corresponding to the dark counts of the detectors was removed from each spectrum. (c) Spatial map of the peak intensity in the 440-480 cm⁻¹ band of 1L-MoS₂. (d) Spatial map of the peak intensity in the 320-380 cm⁻¹ band of the 1L-WS₂ shell. (e) Spatial

map of the peak intensity in the 180-220 cm$^{-1}$ resonant band associated with the alloyed transition region. (f) Combined image of the TERS bands with the 1L-MoS$_2$ band in the red channel, the alloyed transition band in the blue channel, and the 1L-WS$_2$ band in the green channel. Scale bars for (c)-(f) are 400 nm. The peak intensity maps enhance the contrast of the different regions by integrating the intensity of any peaks in the corresponding bands that remain after removing a linear background (see SI for details). The pixel-size in the TERS imaging is 25×25 nm$^2$ and signals were acquired with 250 ms/pixel integration times.

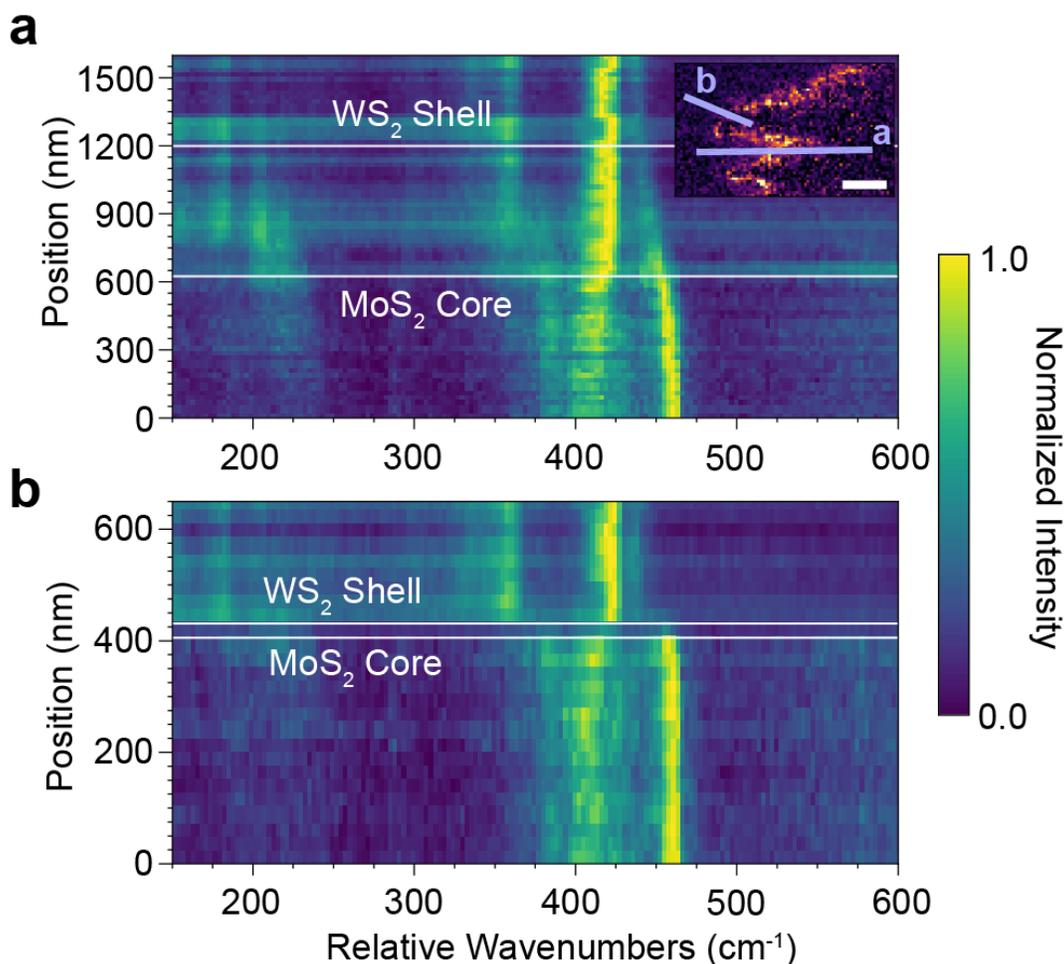

**Figure 5** – characterization of the evolution of the semi-resonant TERS spectra across the broad and sharp transition regions in a 2D lateral heterostructure. (a) The evolution of the semi-resonant TERS spectrum across the broad transition region (path 'a' in the inset). Inset: spatial map of the TERS peak intensity in the spectral region of 180-220 cm$^{-1}$ (duplicated from Figure 4d). Scale bar: 400 nm. The light-blue lines mark the path from which the spectra are interpolated. (b) The evolution of the semi-resonant TERS spectrum across the sharp transition region (path 'b' in the inset). For both (a) and (b), the TERS spectra are interpolated from a 50×25 nm$^2$ region at each position along the respective paths. The beginning and end of the transition regions are estimated as the points where the Raman spectra, specifically the mode at ~450 cm$^{-1}$ in the MoS$_2$ and ~410

cm$^{-1}$ in the WS$_2$, ceases to systematically change with position along the paths. A constant background corresponding to the dark counts of the detectors is removed from each spectrum.

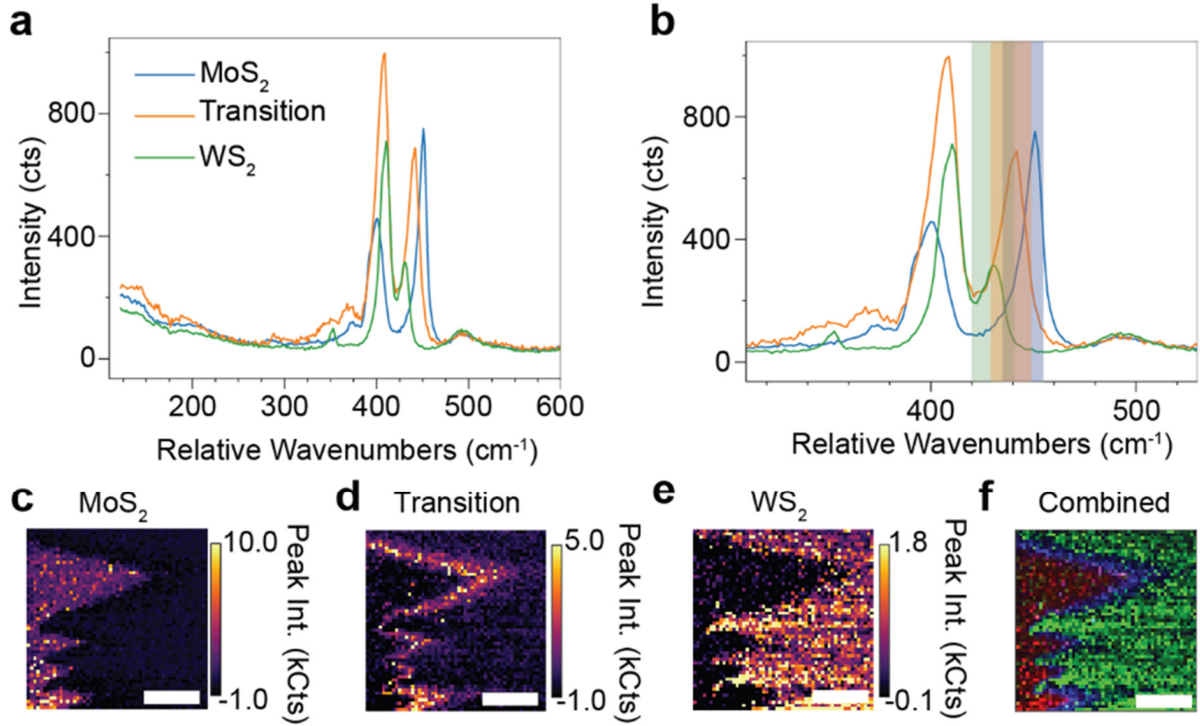

**Figure 6** – non-resonant TERS imaging and spectroscopy of the heterostructure interface. TERS spectrum acquired under non-resonant excitation over (a) 120-600 cm$^{-1}$ and (b) zoomed into the range of 310-530 cm$^{-1}$. The spectra are averages over the following areas: 400 pixels from a 240×960 nm$^2$ area for MoS$_2$, 110 pixels from a 240×264 nm$^2$ for transition, and 400 pixels from a 480×480 nm$^2$ for WS$_2$. A constant background corresponding to the dark counts of the detectors is removed from each spectrum. Peak intensity maps of the (c) high-energy (436-456 cm$^{-1}$; blue band in panel b), (d) intermediate-energy (430-450 cm$^{-1}$; orange band in panel b), and (e) low-energy (420-440 cm$^{-1}$; green band in panel b) bands for the 1L-MoS$_2$ core, alloyed transition region, and 1L-WS$_2$ shell, respectively. (c) Combined intensity map of the images in (b) with the 1L-MoS$_2$ band in the red channel, alloy band in the blue channel, and 1L-WS$_2$ band in the red channel. Scale bars for (c)-(f): 400 nm. The peak intensity maps enhance the contrast of the different regions by integrating the intensity of any peaks in the corresponding bands that remain

28/39

after removing a linear background (see SI for details). The pixel size of the TERS imaging is $24 \times 24$ nm$^2$ and the signals were acquired with 100 ms/pixel integration times.

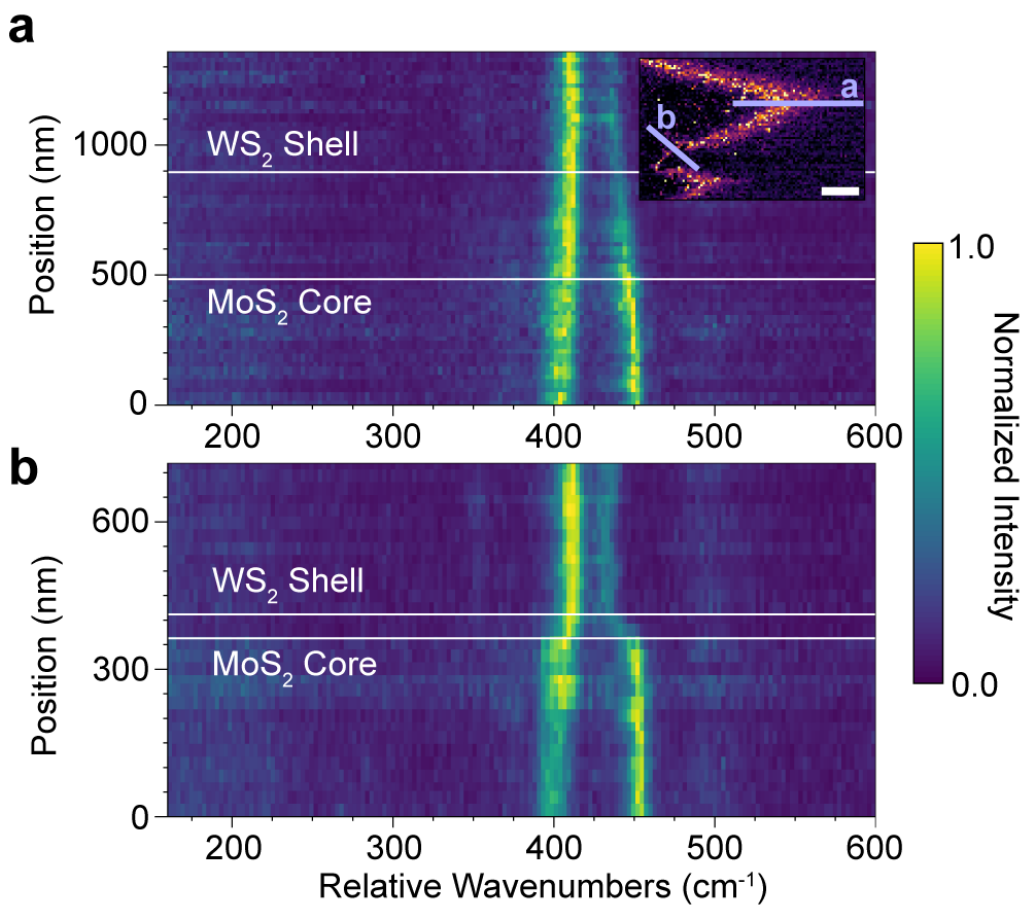

**Figure 7 -** characterization of the broad and sharp transition regions of the interface in a 2D lateral heterostructure using non-resonant TERS. (a) The evolution of the non-resonant TERS spectrum across the broad transition region (path 'a' in the inset). Inset: spatial map of the TERS peak intensity in the spectral region of 430-450 cm⁻¹ (duplicated from Figure 6d). Scale bar: 400 nm. The light-blue lines mark the path from which the spectra are interpolated. (b) The evolution of the non-resonant TERS spectrum across the sharp transition region (path 'b' in the inset). For both (a) and (b), the TERS spectra are interpolated from a $48{\times}24$ nm² region at each position along the respective paths. The beginning and end of the transition regions are estimated as the points where the Raman spectra, specifically the mode at ~450 cm⁻¹ in the $MoS_2$ and ~410 cm⁻¹ in the $WS_2$,

ceases to systematically change with position along the paths. A constant background corresponding to the dark counts of the detectors is removed from each spectrum.

**TOC Graphic**

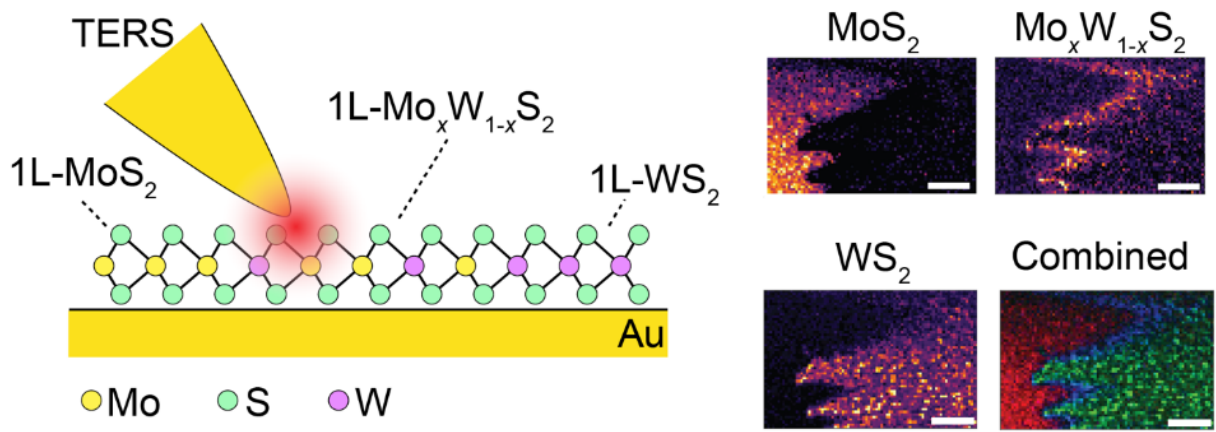

# Nanoscale Raman characterization of a 2D semiconductor

# lateral heterostructure interface

## Supporting Information

*Sourav Garg[1,1], J. Pierce Fix[2,†], Andrey V. Krayev[3], Connor Flanery[2], Michael Colgrove[2],*

*Audrey R. Sulkanen[4], Minyuan Wang[4], Gang-Yu Liu[4], Nicholas J. Borys[2,2], and Patrick Kung[1,3]*

[1]Department of Electrical and Computer Engineering, The University of Alabama,

Tuscaloosa, AL, 35487, USA

[2]Department of Physics, Montana State University, Bozeman, MT, 59717, USA

[3]HORIBA Scientific, Novato, CA, 94949, USA

[4]Department of Chemistry, University of California Davis, Davis, CA, 95616, USA

---

[1] These authors contributed equally

[2] nicholas.borys@montana.edu

[3] patkung@eng.ua.edu

**Section 1: Optical images of jagged exterior edges of the WS₂ shell of the heterostructure crystallites on gold.**

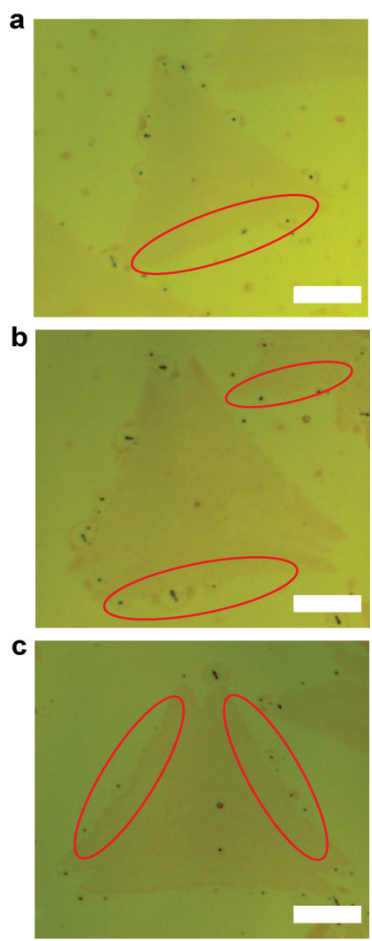

**Figure S1** – (a,b,c) Optical images of heterostructure crystallites on a gold film with jagged exterior edges in the 1L-WS₂ shell. Scale bars in (a), (b) and (c): 10 μm. Red ovals highlight jagged regions.

## Section 2: Transfer of 2D Nanostructues onto Au Thin Films

The sample before and after peeling were imaged using optical microscopy, as shown in Figure S2a and Figure S2b respectively. A cluster of 2D heterostructures shown in Figure S2a (yellow enclosure) appear as bright contrast in optical microscope. The same area was imaged again after peeling. The location where the old cluster region appears darker indicating the successful transfer from the sapphire surface onto the Au thin film.

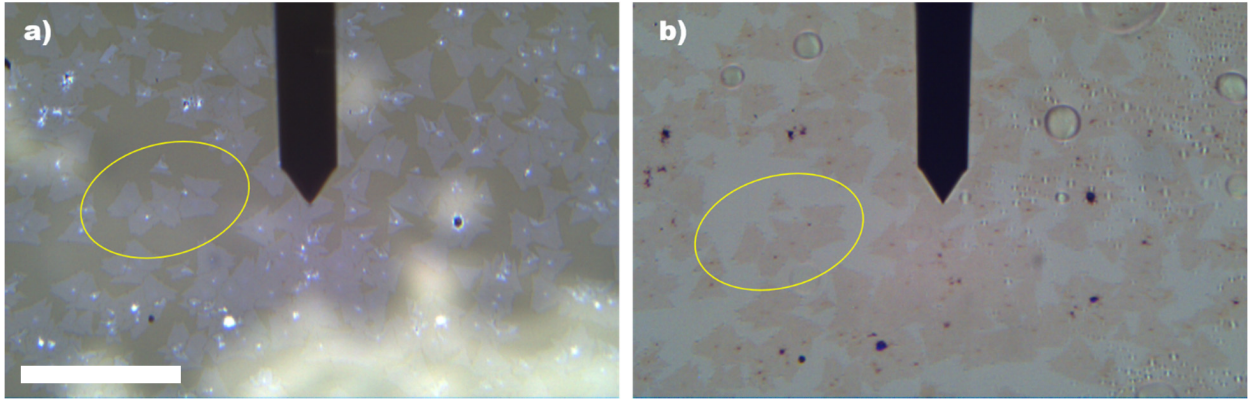

**Figure S2 -** (a) Optical microscopy image of the $MoS_2/WS_2$ crystals on sapphire surface. The top feature in (a) and (b) is a AFM prob. (b) The same area after peeling indicate the location where the 2D nanocrystals were peeled off from the sapphire surface. Scale bars in (a), and (b): 100 µm.

## Section 3: transition widths from prior CPD imaging results

Table S1 summarizes the measured widths of transition regions in 2D lateral heterostructures using CPD imaging that have been reported in the prior studies cited in the main text. The widths were extracted by digitizing the published CPD line traces.

**Table S1:** Summary of transition widths reported in prior CPD imaging measurements.

| Reference in the main text | Figure in reference | Heterostructure | Width (nm) |
|---|---|---|---|
| 10 | 4d | $WS_2/WS_{1.44}Se_{0.56}$ | 1400 |
| | 4g | $WS_2/WS_{1.04}Se_{0.96}$ | 780 |
| | 4i | $WS_2/WSe_2$ | 3160 |
| 22 | 4e | $WSe_2/MoS_2$ (dark) | 550 |
| | 4e | $WSe_2/MoS_2$ (illuminated) | 660 |
| 23 | 3e | $WS_2/MoS_2$ | 1700 |
| 24 | 4b | $WS_2/MoS_2$ | 2760 |

## Section 4: Peak filtering analysis

To make the TERS intensity maps in Figures 4 and 6 in the main manuscript, a peak filtering analysis was used to isolate peaks within the spectral window from background tails from other neighboring peaks. The function fits a linear line to four points that include the two data points with the highest and lowest wavenumbers on the high and low-energy sides of the selected spectral range, respectively (see green line in Figure S3d, e). The four data points are defined by the spectral range analyzed, and the same data points are consistently used to analyze the TERS spectrum and calculate the peak intensity of each pixel. They are not adjusted on a per-pixel basis. The linear fit is then subtracted from the raw data leaving an isolated peak when a peak is clearly in the band. In Figure S3a we have shown the wavenumber range used to map the transition area. The intensity map shown in Figure S3b demonstrates an integrated intensity map over the defined wavenumber range in Figure S3a but without the peak filtering. The transition area in Figure S3b is difficult to see due to interference from overlapping modes in the $MoS_2$ and $WS_2$ in this range. In contrast, Figure S3c shows the intensity map after the peak filtering. In Figures S3c, the intensity from the transition region is isolated with minimal interference from the $MoS_2$ and $WS_2$. In Figures S3d and S3e, we plot the raw averaged spectra in the defined wavenumber range from the transition and $WS_2$ areas respectively before and after the peak filter was applied. These spectra show that we can effectively isolate the ~440 cm$^{-1}$ peak in the transition area so interference from overlapping peaks, like the ~433 cm$^{-1}$ peak in Figure S3e, is minimized.

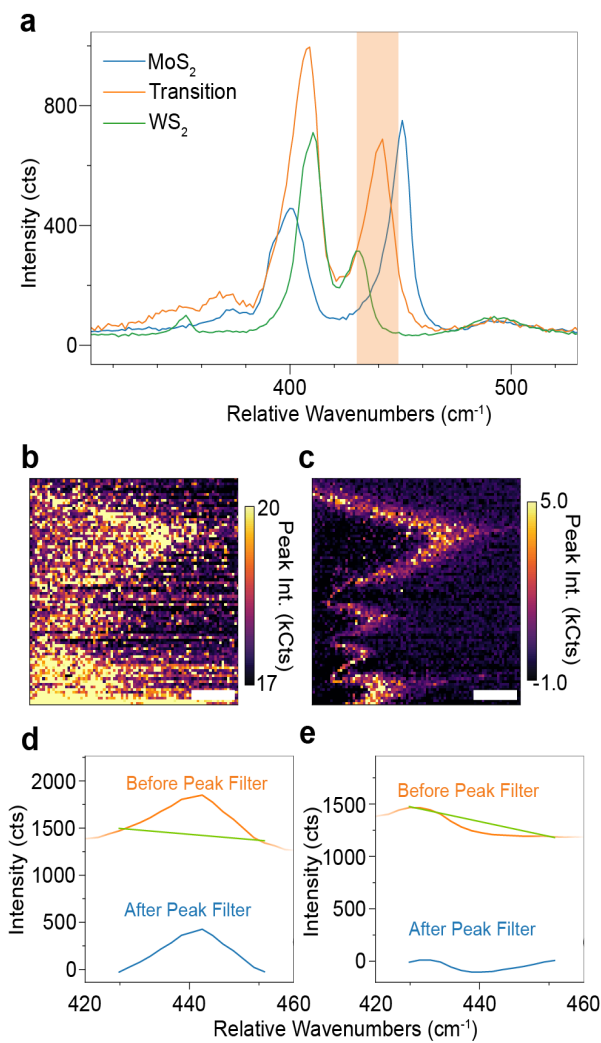

**Figure S3** – Explanation of the peak filtering analysis. (a) TERS spectra acquired under non-resonant excitation with the orange band marking the wavenumber range of interest (430-450 $cm^{-1}$). (b) Integrated intensity map of the wavenumber range defined in panel (a) without the peak filtering analysis. (c) Peak intensity map of the wavenumber range defined in panel (a). Scale bars in (b) and (c): 400 nm. (d) The average TERS spectrum of the transition region before (orange) and after (blue) the peak filtering analysis over the wavenumber range 430-450 $cm^{-1}$. The curves are offset for clarity. (e) The average spectrum for the 1L-$WS_2$ shell before (orange) and after (blue) the peak filtering analysis over the wavenumber range 430-450 $cm^{-1}$. The curves are offset vertically for clarity. In panels (d) and (e), the green lines indicate the linear fits that was were subtracted in the peak filtering analysis.

## Section 5: Alloy composition gradient analysis

To prove that the spectral position of the 442 cm$^{-1}$ peaks in the TERS spectra does depend on the alloying ratio in the $Mo_xW_{1-x}S_2$ compound, we performed a systematic study on the correlation between the position of this peak in TERS spectra of gold-transferred crystals with the position of the PL peak in the same crystals on the growth sapphire substrate. The correlation results were obtained on a gradient composition sample where the PL peak position in $Mo_xW_{1-x}S_2$ crystals varied from 635.5 nm to 658 nm depending on the location of the crystal on the substrate. The spectral position of the composition-sensitive peak in TERS spectra of the same crystals after the gold-assisted transfer consistently changes from 437 cm$^{-1}$ to 444 cm$^{-1}$

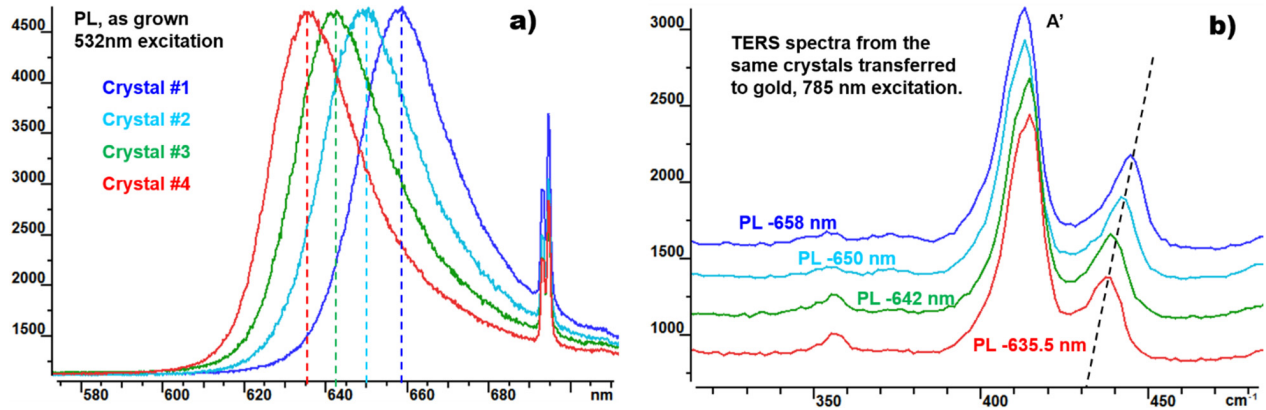

**Figure S4** – (a) Confocal PL spectra from four different $Mo_xW_{(1-x)}S_2$ crystals with varying compositions. The PL spectra that were acquired with 532 nm laser excitation are normalized to the same intensity. (b) Corresponding TERS spectra from the same crystals after the gold-assisted transfer. Spectra are normalized by the intensity of the $A'$ peak and offset vertically for clarity.

(see Figure S4 and Table S1). Thus, we can claim with certainty that the position of the TERS peak in the 430-450 cm$^{-1}$ range depends on the alloying ratio in $Mo_xW_{1-x}S_2$ compounds, though the exact correspondence between the PL peak maximum and the TERS peak position may be different for different samples. For example, varying concentrations of sulfur deficiencies, doping, and/or contributions of charged exciton emissions in different samples may cause such a discrepancy, including inducing changes in the position of the combined PL peak. The composition of each flake (Table S1) is estimated from the PL spectrum using Vegard's Law as discussed in the manuscript.

**Table S2** – Summary of the relation between the position of the PL peak in the as-grown crystals with the position of the composition-sensitive TERS peak after the gold-assisted transfer.

|  | Crystal #1 | Crystal #2 | Crystal #3 | Crystal #4 |
|---|---|---|---|---|
| PL | 658 nm | 650 nm | 642 nm | 635.5 nm |
| TERS | 443.9 cm$^{-1}$ | 441.6 cm$^{-1}$ | 438 cm$^{-1}$ | 437 cm$^{-1}$ |
| Composition | $Mo_{0.29}W_{0.71}S_2$ | $Mo_{0.19}W_{0.81}S_2$ | $Mo_{0.11}W_{0.89}S_2$ | $Mo_{0.05}W_{0.95}S_2$ |

## Section 6: Resonant TERS of single-layer alloys exfoliated from bulk crystals

To find the alloy composition in 1L-$Mo_xW_{1-x}S_2$ at which the 140-240 cm$^{-1}$ Raman modes have the highest intensity we performed resonant TERS measurements of four alloy compositions of 1L-$Mo_xW_{1-x}S_2$, 1L-$MoS_2$, and 1L-$WS_2$ that were mechanically exfoliated from bulk crystals, which ideally have lower amounts of defects. Figure S5 shows the progression of the TERS spectra of these samples from pure $WS_2$ to pure $MoS_2$. The spectra in Figure S5 are averages from a scan area of ~500×500 nm$^2$. A broad PL background was removed from each spectrum using a polynomial fit. We observe the highest intensity in the 140-240 cm$^{-1}$ modes in the $Mo_{0.15}W_{0.85}S_2$ and $Mo_{0.30}W_{0.70}S_2$ alloys. As the tungsten concentration decreases in the alloys, the 140-240 cm$^{-1}$ modes decrease in intensity until we reach pure $MoS_2$ where the modes are weak indicating a possible alloying effect in the material. We note that the high energy mode associated with S vacancies as discussed in the manuscript for 1L-$WS_2$ are not present in the TERS spectra of the exfoliated 1L-$WS_2$ measured here.

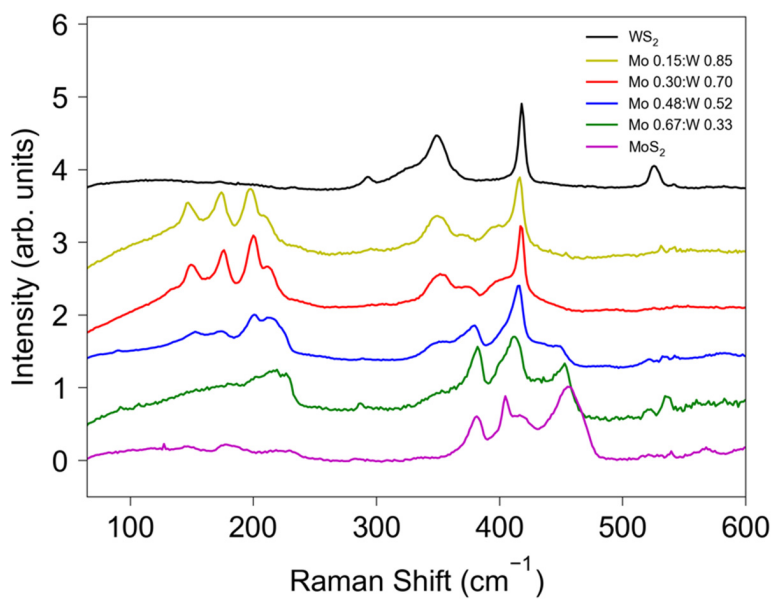

**Figure S5 -** Evolution of the TERS spectra of exfoliated and gold-transferred monolayers of different composition. Appearance of characteristic Raman peaks within 140-240 cm$^{-1}$ range material composition evolves from pure $WS_2$ to alloyed $Mo_xW_{(1-x)}S_2$ compound is clearly seen. The intensity of these Raman bands is maximized at the W/Mo ratio of about 2.32 and then decreases coming to practically zero in pure $MoS_2$.

## Section 7: Topography of region studied by TERS

Figure S6 below presents the detailed characterization of the topography and CPD of the region analyzed in the TERS imaging presented in Figures 4 and 5 in the main manuscript.

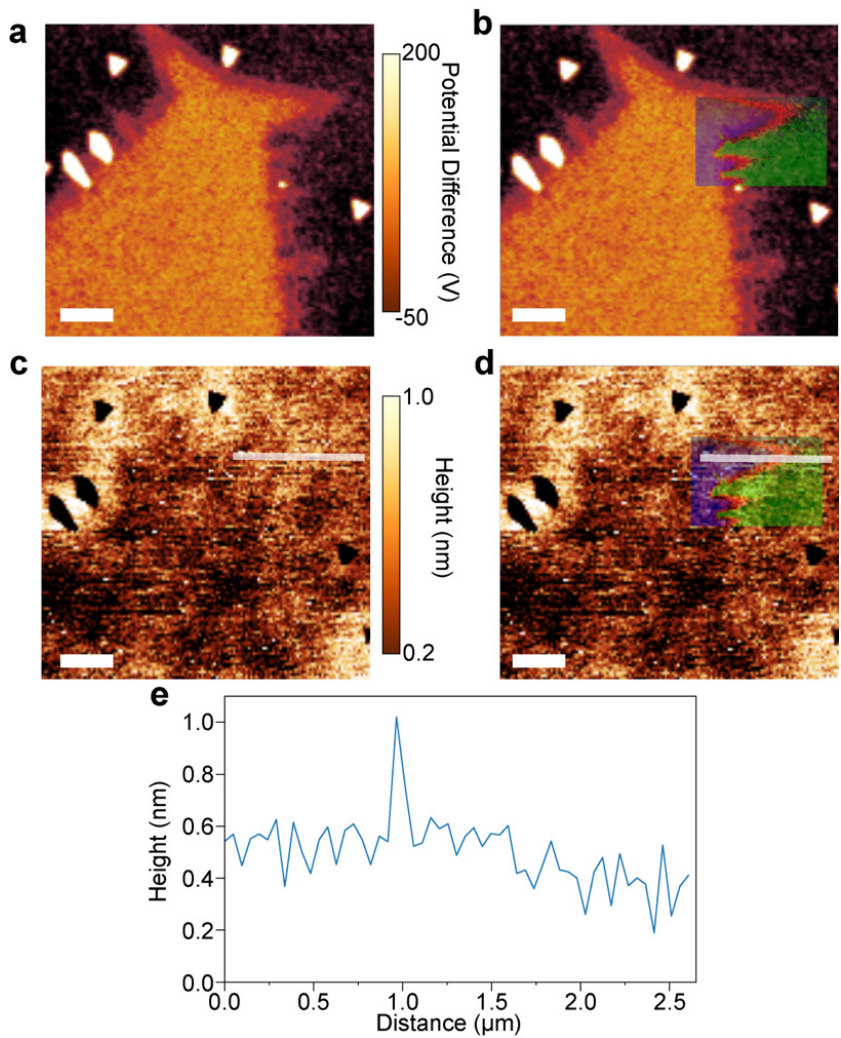

**Figure S6 –** (a) CPD image of the region characterized with TERS. (b) CPD image with TERS image superimposed. (c) AFM image of the region characterized with TERS. (d)AFM image with TERS image superimposed. (e) Topographic profile of the 2D heterostructure along the white line in (c and d). Scale bar: *1 μm*. Superimposed TERS map is of the region measured for Figures 4 and 6.

## Section 8: Extended line traces beyond the interfacial region for the resonant and non-resonant TERS measurements

To demonstrate that the TERS spectrum does not evolve beyond the transition region, we have extended the line traces over the largest extent of the data from Figures 5 and 7. The extended line traces are shown below in Figure S7.

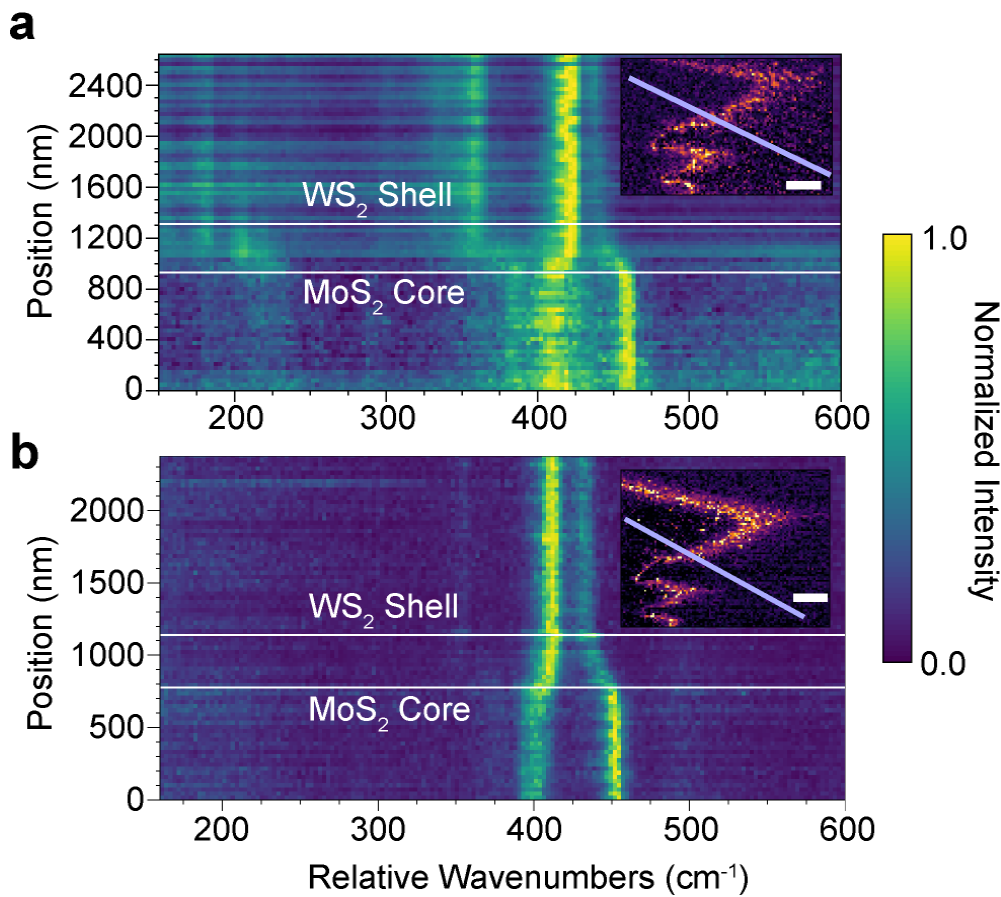

**Figure S7** - characterization of the transition regions of the interface in a 2D lateral heterostructure using resonant and non-resonant TERS. (a) The evolution of the resonant TERS spectrum across the extent of the data (path shown in the inset). Inset: spatial map of the TERS peak intensity in the spectral region of 430-450 cm⁻¹ (duplicated from Figure 6d). Scale bar: 400 nm. The light-blue lines mark the path from which the spectra are interpolated. (b) The evolution of the non-resonant TERS spectrum across the extent of the data (path shown in the inset). For both (a) and (b), the TERS spectra are interpolated from a 48×24 nm² region at each position along the respective paths. The beginning and end of the transition regions are estimated as the points where the Raman spectra, specifically the mode at ~450 cm⁻¹ in the $MoS_2$ and ~410 cm⁻¹ in the $WS_2$, ceases to systematically change with position along the paths. A constant background corresponding to the dark counts of the detectors was removed from each spectrum.